\documentclass[%
 aip,
 jmp,%
 amsmath,amssymb,
 reprint,%
]{revtex4-1}

\usepackage{graphicx}
\usepackage{dcolumn}
\usepackage{bm}

\begin{document}

\title{Impedance of rigid bodies in one-dimensional elastic collisions}

\author{Janilo Santos}
\email{janilo@dfte.ufrn.br}
\affiliation{Departamento de F\'{\i}sica, Universidade Federal do Rio G. do Norte,  59072-970 Natal - RN, Brazil}
\author{Bruna P. W. de Oliveira}
\email[]{deolivei@usc.edu}
\affiliation{Department of Physics and Astronomy, University of Southern California, Los Angeles, CA 90089}
\author{Osman Rosso Nelson}
\email[]{osman@dfte.ufrn.br}
\affiliation{Departamento de F\'{\i}sica, Universidade Federal do Rio G. do Norte,  59072-970 Natal - RN, Brazil}
\date{\today}%

\begin{abstract}
In this work we study the problem of one-dimensional elastic collisions of billiard balls, considered as rigid bodies, in a framework very different from the classical one presented in text books. Implementing the notion of impedance matching as a way to understand efficiency of energy transmission in elastic collisions, we find a solution which frames the problem in terms of this conception. We show that the mass of the ball can be seen as a measure of its impedance and verify that the problem of maximum energy transfer in elastic collisions can be thought of as a problem of impedance matching between different media. This approach extends the concept of impedance, usually associated with oscillatory systems, to system of rigid bodies.
\end{abstract}
\maketitle

\section{Introduction}

A good teacher knows the value of analogy and universality when explaining difficult concepts. A hard problem can be much simpler to elucidate when students have been exposed to a similar problem. For instance, students understand electrical forces better after they are acquainted with gravitational forces. Terms such as \textit{energy} become gradually more familiar as it is encountered in a variety of contexts. In this paper we aim to introduce and enlarge the concept of \textit{impedance} to undergraduates and advanced high school students by investigating energy transfer in mechanical collisions and tracing a parallel with the propagation of light in electromagnetic systems. We define the \textit{characteristic impedance} of a system as the ratio between a \textit{force-like quantity} and \textit{a velocity-like quantity} \cite{Crawford}. From this definition we derive an expression for the mechanical impedance of a billiard ball, which tells how to enhance the energy transfer from one mass to another in elastic collisions. Most importantly, we investigate how impedance matching appears in mechanical systems and we compare our results with the well-known problem of impedance matching in optical systems.

\section{Transmission of kinetic energy in a head-on elastic collision}

Our mechanical system consists of the one-dimensional elastic non-relativistic collision between two or three particles with different masses. This is simply the popular textbook problem of one-dimensional elastically colliding billiard balls. We observe how much kinetic energy is transmitted from one ball to the other during the collision.

\begin{figure*}[t]
	\includegraphics[scale=0.6] {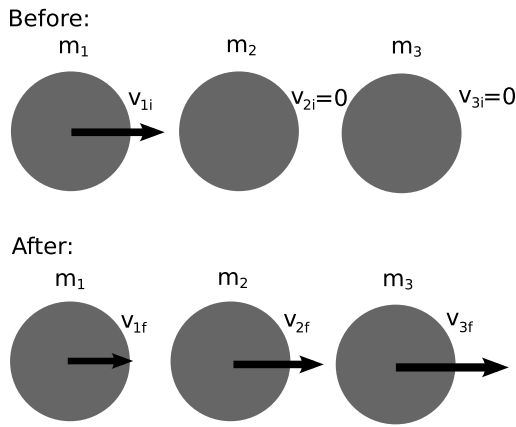}
	\caption{\small Physical picture of the one-dimensional elastic collision between balls of masses $m_1$, $m_2$ and $m_3$. Before the collisions, ball 1 has speed $v_{1i}$ and balls 2 and 3 are at rest. After the collisions, all balls have different speeds if their masses are different.
		}
\label{f1}
\end{figure*}

\begin{figure*}[t]
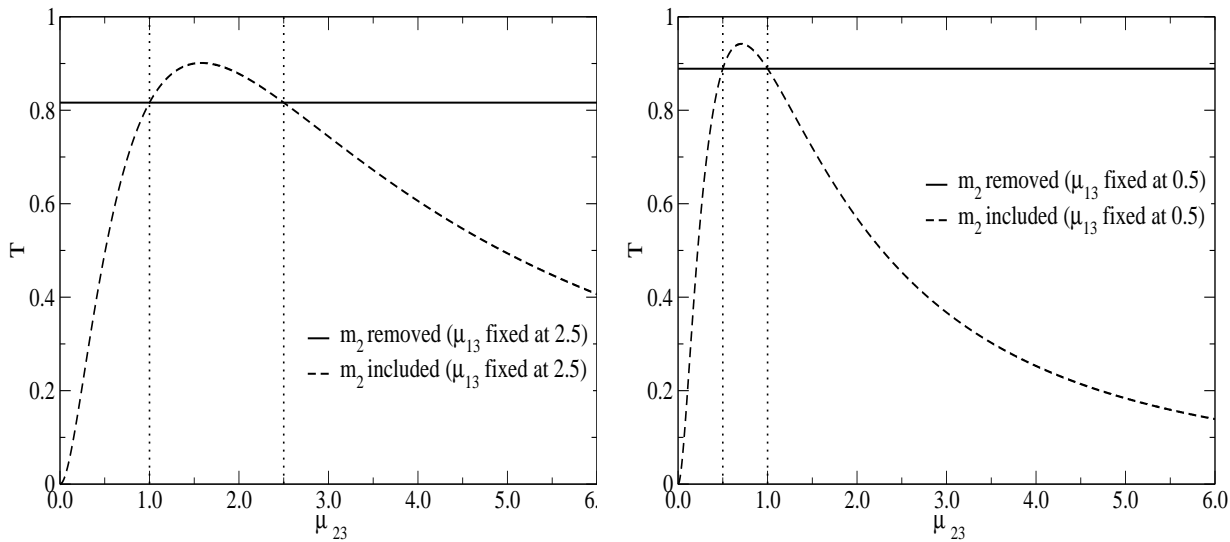

\flushleft{\includegraphics[width=3.15truein,height=2.8truein]{101122-1}}
\hspace{0.00in}
\includegraphics[width=3.15truein,height=2.8truein]{101122-2}
\caption{\small Plots of Eq. (\ref{T_transferred}) for fixed $\mu_{13}>1$ (left) and for $\mu_{13}<1$ (right). We notice that in the presence of $m_2$, there is a range of values of $\mu_{23}$ (limited by the vertical lines) where the fraction of energy transmitted is greater than in its absence. The regions are limited by the values of $\mu_{13}$ and $\mu_{23}=1$.}
\label{intermediate_masses}
\end{figure*}

Before we introduce the idea of impedance in mechanical systems, let us use the conservation laws of linear momentum and kinetic energy in elastic collisions to find the fraction of transmitted energy from one object to the other. We consider three rigid billiard balls of masses $m_1$, $m_2$, and $m_3$. Let us assume that ball 1 has a finite speed and both balls 2 and 3 are at rest before the collision. After the first collision between $m_1$ and $m_2$, part of the kinetic energy from ball 1 has been transmitted to ball 2, which now has a velocity in the same direction as the initial velocity of ball 1 (see Fig.~\ref{f1}). Using momentum and kinetic energy conservation, one obtain, for the fraction of kinetic energy transmitted from the first to the second ball,
\begin{equation} \label{T_12}
T_{12}=\frac{K_{2f}}{K_{1i}}=\frac{4\mu_{12}}{(1+\mu_{12})^2},
\end{equation}
where we define $\mu_{12}\equiv m_1/m_2$, $K_{1i}$ is the initial kinetic energy of ball 1 and $K_{2f}$ is the kinetic energy of ball 2 after the collision. The fraction of energy that remains in the first ball, which we consider as a ``reflected" energy, is given by
$R_{12}=K_{1f}/K_{1i}=(\mu_{12}-1)^2/(\mu_{12}+1)^2$, where $K_{1f}$ is the kinetic energy of the first ball after the collision. Analogously, in the second collision between balls 2 and 3, the fraction of kinetic energy that is transferred to the third ball is
\begin{equation} \label{T_23}
T_{23}=\frac{K_{3f}}{K_{2f}}=\frac{4\mu_{23}}{(1+\mu_{23})^2},
\end{equation}
where $\mu_{23}\equiv m_2/m_3$ and $K_{3f}$ is the kinetic energy of $m_3$ after the second collision. The fraction of energy transferred from the first to the third ball in the process is given by
$T_{13}=K_{3f}/K_{1i}=T_{12}T_{23}$, which can be written, using (\ref{T_12}) and (\ref{T_23}), as
\begin{equation} \label{T_transferred}
T_{13}=\frac{16\mu_{13}}{\left(1+\mu_{13}+\frac{\mu_{13}}{\mu_{23}}+\mu_{23}\right)^2},
\end{equation}
where we define $\mu_{13}\equiv m_1/m_3$ and $\mu_{12}$ has been replaced with the equivalent expression $\mu_{13}/\mu_{23}$.  From this equation we see that, for any fixed value of $\mu_{13}$, there are many values of $\mu_{23}$ which give different fractions of transmitted kinetic energy from the first to the third ball. We compare it with the configuration when the intermediate
ball $m_2$ is removed, in which case the transferred energy is given by $T_{13}=4\mu_{13}/(1+\mu_{13})^2$. Equating this with Eq. \ref{T_transferred} we find two roots: $\mu_{23}=1$ and $\mu_{23}=\mu_{13}$. The plot for this configuration is shown in Fig.~\ref{intermediate_masses}, where we examine
the behavior of (\ref{T_transferred}) for two particular values of the ratio $\mu_{13}$. We observe that, for each $\mu_{13}$, there exists a range of values between $\mu_{23}=1$ and $\mu_{13}$, such that more energy is transmitted in the presence of $m_2$ than when this intermediate mass is absent.


\begin{figure*}[t]
\centering
\includegraphics[scale=0.4]{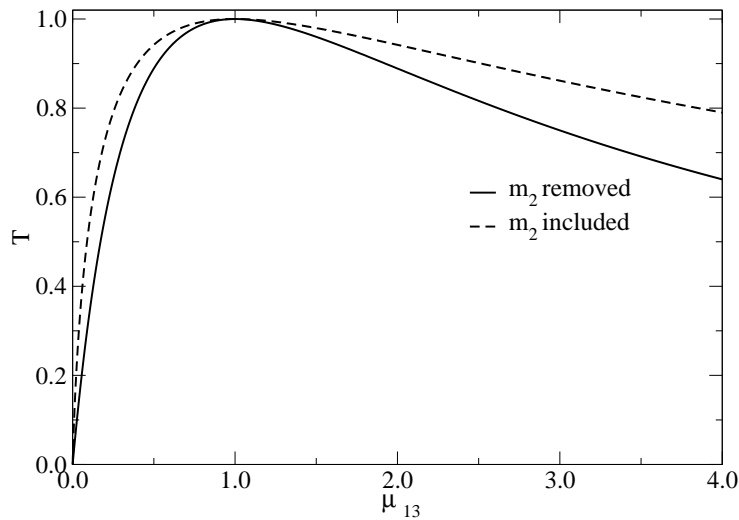}
\vspace{0cm}
\caption{\small Kinetic energy transmission coefficient (Eq.~\ref{T_max}) from ball $m_1$ to ball $m_3$ in Fig.~\ref{f1}, when the intermediate ball has mass $m_2=\sqrt{m_1m_3}$.  Note that the transmission with $m_2$ included (dashed curve) is always greater then the other when $m_2$ is removed (solid curve).}
\label{f2}
\end{figure*}

In order to proceed further, we ask ourselves whether this special range of values can be enlarged such that a maximum amount of kinetic energy can be transferred from the first to the third ball. Indeed, fixing $m_1$ and $m_3$, this can be obtained by taking $dT_{13}/d\mu_{23}=0$. Investigating the second derivative, we find that $T_{13}$ has a maximum at $\mu_{23}=\sqrt{\mu_{13}}$, that is, when we take $m_2=\sqrt{m_1m_3}$. The fraction of transferred energy in this case depends only on the ratio $\mu_{13}$ and is given by
\begin{equation}  \label{T_max}
T_{13max}=\frac{16\mu_{13}}{(1+\mu_{13}+2\sqrt{\mu_{13}})^2}\,.
\end{equation}
This answers our question about the value of the intermediate mass: when $m_2$ is equal to the geometric mean of $m_1$ and $m_3$, the transmitted kinetic energy is a maximum. Fig.~\ref{f2} shows the behavior of Eq.~(\ref{T_max}) for several values of the ratio $\mu_{13}$ and compares it with the configuration where the intermediate ball $m_2$ is absent. We observe that with an intermediate ball with mass $m_2=\sqrt{m_1m_3}$ we always transfer more kinetic energy from $m_1$ to $m_3$, and when $\mu_{13}=1$, that is, $m_1=m_3$, the presence of $m_2$ is irrelevant (transmission coefficient is equal to unity). A similar calculation for partially elastic collisions between $n$ masses was carried out by J. B. Hart and R. B. Herrmann \cite{Hart}. We expand on their results by emphasizing the analogy with impedance matching in the following sections.

If this is done in class as a demonstration, the students will be faced with the question that arises from the results: \textit{Why does the presence of the intermediate ball facilitate the transmission of energy? Wouldn't it be more reasonable to expect that the presence of an extra ball would reduce the transmission of kinetic energy}?
This question, as we will see in the following sections, is more easily answered if it is introduced in the context of \textit{impedance matching}.

\section{Impedance matching}

We know from electromagnetism that the transfer of energy through the interface between different media depends on their respective values of impedance $Z$. For an electromagnetic wave traveling from, say, medium 1 to medium 3, the coefficients of reflection ($r_{13}$) and transmission ($t_{13}$), known as Fresnel coefficients~\cite{Reitz}, are associated with the fraction of reflected energy $R_{13}$ and transmitted energy $T_{13}$, such that $R_{13}=r_{13}^2$, $T_{13}=(n_3/n_1)t_{13}^2$ ($n_1$ and $n_3$ are the indices of refraction of the media 1 and 3 respectively and $r_{13}=1-t_{13}$). Although in optical systems the coefficient $r_{13}$ is given in terms of the indices of refraction as  $r_{13}=(n_3-n_1)/(n_3+n_1)$, more generally it can be expressed in terms of the impedances of the media as $r_{13}=(Z_3-Z_1)/(Z_3+Z_1)$, where $Z_i$ is the impedance of medium $i$. Since the sum of the reflected and transmitted parts has to be unity, we obtain $t_{13}=2Z_1/(Z_3+Z_1)$.
Therefore, when the two media have the same impedance, all energy is transmitted and $t_{13}=1$, $r_{13}=0$. This problem is similar to the mechanical problem we described in the previous section if we add an intermediate medium with impedance $Z_2$. Once again, we are interested in the energy transfer in the problem and we can ask the question: \textit{What is the value of $Z_2$ for which the transmission of energy from medium 1 to medium 3 is maximum}?

In order to solve this problem, we note that in this configuration the fraction of energy transmitted from medium 1 to medium 2 is $T_{12}=4Z_1Z_2/(Z_1+Z_2)^2$ and the fraction transmitted from medium 2 to medium 3 is $T_{23}=4Z_2Z_3/(Z_2+Z_3)^2$. Thus, the transmission coefficient from medium 1 to medium 3, $T_{13}=T_{12}T_{23}$, is given by
\begin{equation}
T_{13}=\frac{16Z_1Z_2^2Z_3}{(Z_1+Z_2)^2(Z_2+Z_3)^2}.
\end{equation}
The maximum transmission $(dT_{13}/dZ_2 =0)$ occurs for $Z_2=\sqrt{Z_1Z_3}$. The value of $Z_2$ that allows for maximum energy transfer from medium 1 to  medium 3 is the geometric mean of $Z_1$ and $Z_3$, which represents the so-called \textit{impedance matching}. This derivation can be found in many advanced textbooks in electromagnetism, acoustics and optics~\cite{Graham}; in introductory texts of physics usually impedance matching is only briefly mentioned in the study of electric circuits \cite{Young, Halliday}. It is worth mentioning that impedance matching is also the concept behind the anti-reflective coatings found in eyeglasses, binoculars, and other lenses. Notice that the value found for the matching impedance $Z_2$ resembles our previous result for the intermediate mass $m_2$ in section II ($m_2=\sqrt{m_1m_3}$). We explore this in the next section.

\section{Impedance of a rigid billiard ball}

We now return to the concept of impedance as the ratio between a force-like quantity and a velocity-like quantity \cite{Crawford} in order to find out what would play the role of impedance in a mechanical system such as rigid billiard balls.  As investigated in Section II, in these collisions the force-like quantity is not simply the force $\vec F$ due to the collision, but the integrated effect of this force during the collision time $\Delta t= t_f - t_i$. This is the impulse $\vec J$ that the target ball receives from the incident one. Here, $t_i$ and $t_f$ are the initial and final time of collision, respectively. Therefore, in the general case of a frontal collision between two balls in which the target ball is at rest, we obtain
\begin{equation}  \label{Impulse}
\vec{J}=\int_{t_i}^{t_f}\vec{F}(t)dt=\vec{p_f}-\vec{p_i}=m\vec{v_f}\,.
\end{equation}
In Eq. (\ref{Impulse}) $\vec v_f$ is the response of the ball to the impulse $\vec  J$. We ascribe the impedance
\begin{equation}
Z=\frac{J}{v_f}=m
\end{equation}
to a rigid billiard ball, considered as a particle of mass $m$. This explains why the presence of the intermediate ball facilitates the transmission of energy in the elastic collisions studied in Sec. II. The choice $m_2=\sqrt{m_1m_3}$ works as an impedance matching between two media of impedances $Z_1=m_1$ and $Z_3=m_3$.

\section{Conclusion}

In a very well-known problem in classical mechanics, one aligns three rigid balls of different masses $m_1$, $m_2$, and $m_3$. The value of $m_2$, a function of $m_1$ and $m_3$, is to be determined such that when the one-dimensional collisions between these objects are elastic, the transmission of kinetic energy from the first ball to the last ball is maximized. This problem is easily solved using the laws of energy and linear momentum conservation, and we verify that the presence of an intermediate ball enhances rather than suppresses the transmission of energy. In this paper we present an explanation for this problem by proposing an extension of the concept of impedance, usually associated with oscillatory systems, to a rigid billiard ball. We have shown that in the case of one-dimensional elastic collisions, the mass of a particle can be seen as a measure of its impedance. Once this is assumed, we verify that for maximum energy transfer the intermediate mass must be chosen such that it matches the impedances of the first and third mass, each considered as a different medium with their respective impedances. This can be easily explored in the classroom, either by a computer simulation or an actual experiment (an experimental device has been proposed by Hart and Herrmann~\cite{Hart}). Once students are exposed to the idea behind impedance matching with a simple classical collision problem, this can be expanded into a discussion of impedance of electric circuits, acoustics, and optical media.

\section{Acknowledgments}
Author Bruna P.W. Oliveira would like to thank N.T. Jacobson for his useful comments and review of the manuscript. J. Santos thanks the financial support from CNPq and also thanks Prof. M\'ario K. Takeya for helpful discussions about some ideas presented in this article.

\end{document}